\documentclass[aip,
 amsmath,amssymb,
 reprint,%
]{revtex4-1}

\usepackage{graphicx}
\usepackage{dcolumn}
\usepackage{bm}

\usepackage[utf8]{inputenc}
\usepackage[T1]{fontenc}
\usepackage{mathptmx}
\usepackage{etoolbox}

\makeatletter
\def\@email#1#2{%
 \endgroup
 \patchcmd{\titleblock@produce}
  {\frontmatter@RRAPformat}
  {\frontmatter@RRAPformat{\produce@RRAP{*#1\href{mailto:#2}{#2}}}\frontmatter@RRAPformat}
  {}{}
}%
\makeatother
\begin{document}

\preprint{AIP/123-QED}

\title{Proposal for an electrically controlled terahertz micro-oscillator}
\author{V.G. Bordo}
\affiliation{Independent Researcher, 6400 S{\o}nderborg, Denmark.}
  \email{vgbordo@gmail.com}

\date{\today}

\begin{abstract}
A source which generates sustained terahertz electromagnetic radiation (oscillator) controllable by an applied voltage is proposed. The structure consists of a nanocomposite slab containing metal nanorods enclosed between two graphene sheets which are electrostatically gated. The principle of its operation is based on a positive feedback which the nanorods polarization experiences from the radiation reflected by the graphene mirrors. The proposed approach can provide a microscopic source of monochromatic polarized terahertz radiation of power sufficient for biomedical and security applications.
\end{abstract}

\maketitle

Terahertz radiation is in high demand in diverse applications, including biomedicine, communications and security. Nowadays there is a variety of methods to generate radiation in the terahertz spectral range, such as globars, plasma sources, gyrotrons and synchrotrons, gas and semiconductor lasers, diodes and others.\cite{Sources} Terahertz pulses can also be generated using photoconductive emitters excited by femtosecond lasers.\cite{Heinz04} Recently, continuous-wave lasing pumped by a quantum cascade laser which can be tuned over a wide terahertz range has been demonstrated as well.\cite{Capasso22}\\
Despite a long-term development, the available terahertz sources either suffer from broad emission spectrum and low output power or they are so complex and bulky that cannot be used in portable devices. The latter disadvantage is especially critical in biomedical and security applications where such a source should be compatible with the lab-on-a-chip technology.\cite{Menikh02,Lu06}\\
In the present paper, we propose a principally new approach to the generation of terahertz radiation. It is based essentially on the same effect which a dipole antenna experiences above the Earth's surface.\cite{Sommerfeld} The radiation emitted by the antenna is reflected back to it so that the forward and backward pathways interfere and form a feedback loop. The feedback can be either negative or positive, depending on the dipole position. If there is en ensemble of dipoles located nearby, the dipole feels not only its own reflected field, but also the field of all other dipoles. In case where the whole ensemble is confined within the wavelength from the surface, all feedback loops are in phase with each other that can lead to its self-excitation accompanied by the generation of electromagnetic waves.\cite{Bordo16,Bordo18}\\
In the terahertz frequency range, the role of the dipole antennas can be accomplished by metal nanorods embedded in a dielectric.\cite{Bordo23} Such a structure might be prepared using the technology which has been exploited for the fabrication of nanocomposite capacitors where a polymer film is loaded with aligned nanowires.\cite{Tang12} This approach being combined with extraordinary properties of graphene, whose reflectivity is controllable through electrostatic gating,\cite{Crozier18} would provide a microscopic oscillator in which the generation threshold is determined by the applied voltage.\\
In this paper, we consider as a model system a dielectric slab of thickness $d$ and dielectric constant $\epsilon$ loaded with metal nanorods and enclosed between two electrostatically gated graphene sheets. For the sake of simplicity, we assume that all nanorods are aligned along one direction, say the $x$ axis, which is parallel to the slab sides, however, all results can be modified for their random orientations, if necessary. We direct the $z$ axis of the coordinate system perpendicularly to the slab. We analyze the evolution of the electromagnetic field in the structure when it is driven by an external field of arbitrarily small amplitude.\\ 
Let us assume that at the moment $t=0$ an external electromagnetic field of frequency $\omega$ and amplitude $E_0$ directed along the $x$ axis is switched on. We extract the time dependence with the driving field frequency in both the local (microscopic) field and the nanorods polarization and represent them as $E(t)=\tilde{E}(t)\exp(-i\omega t)$ and $P(t)=\tilde{P}(t)\exp(-i\omega t)$, respectively. In the Drude model the polarization satisfies the equation\cite{Bordo23}
\begin{equation}\label{eq:Drude}
\frac{d\tilde{P}(t)}{dt}-i\omega\tilde{P}(t)+\frac{\tilde{P}(t)}{\tau}=\frac{i\sigma_m f}{\omega\tau}\tilde{E}(t),
\end{equation}
where $\sigma_m$ is the metal conductivity for a DC current, $\tau$ is the mean time between successive electron collisions in the metal and $f$ is the volume fraction of metal nanorods in the nanocomposite. The local field on the right-hand side of Eq. (1) obeys in its turn an integral equation which accounts not only the field from all polarized nanorods, but also the contribution from the dipole fields reflected by the graphene sheets:
\begin{equation}
    \tilde{E}({\bf r},t)=E_0({\bf r})+\frac{4\pi}{3\epsilon}\tilde{P}({\bf r},t)+\int_{\mathcal{V}} F({\bf r},{\bf r}^{\prime})\tilde{P}({\bf r}^{\prime},t)d{\bf r}^{\prime},
\end{equation}
where the kernel $F({\bf r},{\bf r}^{\prime})$ relates the $x$-component of the reflected dipole field at the point ${\bf r}=(\bm{\rho},z)$ with the $x$-oriented dipole moment located at the point ${\bf r}^{\prime}=(\bm{\rho}^{\prime},z^{\prime})$ and the integration runs over the nanocomposite volume $\mathcal{V}$.\\
The function $F({\bf r},{\bf r}^{\prime})$ can be written in the form of the 2D Fourier integral as
\begin{equation}
    F({\bf r},{\bf r}^{\prime})=\int\frac{d\bm{\kappa}}{(2\pi)^2}f(\bm{\kappa};z,z^{\prime})\exp[i\bm{\kappa}(\bm{\rho}-\bm{\rho}^{\prime})].
\end{equation}
Representing the local field and the polarization as the Fourier integrals as well, one obtains a relation between their Fourier transforms $e(\bm{\kappa};z,t)$ and $p(\bm{\kappa};z,t)$, respectively, as follows
\begin{eqnarray}\label{eq:e}
e(\bm{\kappa};z,t)=e_0(\bm{\kappa};z)+\frac{4\pi}{3\epsilon}p(\bm{\kappa};z,t)\nonumber\\
+\int_{-d/2}^{d/2}f(\bm{\kappa};z,z^{\prime})p(\bm{\kappa};z^{\prime},t)dz^{\prime},
\end{eqnarray}
where $e_0(\bm{\kappa};z)$ is the Fourier transform of the external field. Besides that, the Fourier transform of Eq. (\ref{eq:Drude}) leads to the equation
\begin{equation}\label{eq:p}
    \frac{d p(\bm{\kappa};z,t)}{dt}-(i\omega-\tau^{-1})p(\bm{\kappa};z,t)=i\eta e(\bm{\kappa};z,t)
\end{equation} 
with 
\begin{equation}\label{eq:eta}
 \eta(\omega)=\frac{\sigma_mf}{\omega\tau}   
\end{equation}
Equations (\ref{eq:e}) and (\ref{eq:p}) determine together an integro-differential equation which governs the evolution of the electromagnetic field in the structure under consideration.\\
The Fourier integral of the microscopic field can be regarded as a superposition of plane waves specified by their wave vector component $\bm{\kappa}$ parallel to the slab sides. In what follows, we will be interested in the radiation emitted perpendicularly to the slab which corresponds to $\bm{\kappa}=0$. For the sake of brevity, we shall omit everywhere this argument.\\ 
The quantity $f(z,z^{\prime})$ involves the reflection coefficient for a graphene sheet, which in its turn is determined by the graphene conductivity. In the terahertz frequency domain the conductivity of graphene, $\sigma_g$, originates from the intraband transitions.\cite{Heinz12} For the chemical potential $\mu$ satisfying the condition $\mid\mu\mid\gg k_BT$ with $k_B$ the Boltzmann constant and $T$ the graphene temperature one has\cite{Falkovsky08}
\begin{equation}
   \sigma_g(\omega)=i\frac{4\mid\mu\mid}{\pi\hbar\omega}\sigma_0,
\end{equation}
where $\sigma_0=e^2/(4\hbar)$ with $e$ the elementary charge and $\hbar$ the Planck's constant is the "universal" conductivity of pristine graphene.\cite{Heinz12} We have neglected here the electron-disorder scattering rate in comparison with $\omega$ that does not reduce the generality of the consideration, but simplifies the further analysis. As far as $\mid\sigma_g\mid$ is a small quantity compared with the speed of light in vacuum, $c$, the reflection coefficient can be approximated as\cite{Falkovsky08}
\begin{equation}\label{eq:r}
    r(\omega)\approx -\frac{2\pi\sigma_g(\omega)}{c}.
\end{equation}
Then, in the linear approximation with respect to $\sigma_g$, the function $f(z,z^{\prime})$ is evaluated as follows\cite{Nha96}
\begin{eqnarray}\label{eq:f}
    f(z,z^{\prime})\approx i\frac{2\pi q(\omega)r(\omega)}{\epsilon}\exp\left[iq(\omega)d\right]\nonumber\\
    \times\left\{\exp\left[iq(\omega)(z+z^{\prime})\right]+\exp\left[-iq(\omega)(z+z^{\prime})\right]\right\}
\end{eqnarray}
with $q(\omega)=(\omega/c)\epsilon^{1/2}$.\\
The further analysis can be conveniently performed with the use of the Laplace transform in time
\begin{eqnarray}
    p(\bm{\kappa},s;z)=\int_0^{\infty}p(\bm{\kappa};z,t)e^{-st}dt,
\end{eqnarray}
Assuming the initial condition $P(t=0)=0$, one obtains from Eqs. (\ref{eq:e}), (\ref{eq:p}) and (\ref{eq:f}) an integral equation
\begin{eqnarray}
    (s-i\omega^{\prime}+\tau^{-1})p(s;z)+A(\omega)\exp[iq(\omega)d]\nonumber\\
    \times\left\{\exp[iq(\omega)z]p_+(s)+\exp[-iq(\omega)z]p_-(s)\right\}\nonumber\\
    =\frac{i\eta(\omega)}{s} e_0(z),
\end{eqnarray}
where 
\begin{eqnarray}
 \omega^{\prime}=\omega+\frac{4\pi}{3\epsilon}\eta(\omega), \\
 A(\omega)=\frac{2\pi}{\epsilon}\eta(\omega) q(\omega)r(\omega),\label{eq:A}
 \end{eqnarray}
 and
 \begin{equation}
     p_{\pm}(s)=\int_{-d/2}^{d/2}\exp[\pm iq(\omega)z]p(s,z)dz.
 \end{equation}
Finally, one comes to the vector equation for the vector
\begin{equation}
    \vec{\mathcal{P}}(s)=
    \left(\begin{array}{c}
p_-(s)\\
p_+(s)
\end{array}\right)
\end{equation}
as follows
\begin{equation}\label{eq:vector}
    \left[\hat{M}-\chi(s)\hat{I}\right]\vec{\mathcal{P}}(s)=\frac{B}{s}\vec{\mathcal{E}}_0
\end{equation}
where 
\begin{equation}
    \hat{M}=
    e^{iqd}\begin{pmatrix}
\frac{\sin(qd)}{qd} & 1\\
1 & \frac{\sin(qd)}{qd}
\end{pmatrix},
\end{equation}
\begin{equation}
    \chi(s)=\frac{s-i\omega^{\prime}+\tau^{-1}}{dA},
\end{equation}
$\hat{I}$ is a unit $2\times 2$ matrix, $B=i\eta/(dA)$ and the vector $\vec{\mathcal{E}}_0$ is defined analogously to $\vec{\mathcal{P}}$ in terms of the external field Fourier transform $e_0(z)$.\\
The time evolution of both the nanorods polarization and the electromagnetic field in the structure is determined by the poles of the Laplace transform $p(s;z)$, $s_{\pm}$. The poles in their turn are found from the equations $\chi(s_{\pm})=\chi_{\pm}$ with $\chi_{\pm}$ being the eigenvalues of the matrix $\hat{M}$. If at least one pole lies in the right half-plane of the complex plane of $s$, i.e. $\text{Re}(s_{\pm})>0$, the solution for the electromagnetic field amplitude in the system grows exponentially with time that signifies a self-excitation process and the beginning of the electromagnetic field generation.\cite{Jenkins} This process develops independently on the driving field amplitude $E_0$ which can be arbitrarily small.\\
The eigenvalues of the matrix $\hat{M}$ are found straightforwardly as
\begin{equation}
    \chi_{\pm}=\exp(iqd)\left[\frac{\sin(qd)}{qd}\pm 1\right].
\end{equation}
From here one obtains the poles of the Laplace transform as follows
\begin{equation}
    s_{\pm}=-\tau^{-1}+i\omega^{\prime}+dA\exp(iqd)\left[\frac{\sin(qd)}{qd}\pm 1\right].
\end{equation}
Accordingly, the criterion of the field generation reads as
\begin{equation}
   d\left[\frac{\sin(qd)}{qd}\pm 1\right]\text{Re}\left[A\exp(iqd)\right]>\tau^{-1},
\end{equation}
which has a clear physical sense: the self-excitation rate of the nanorods polarization given by the left-hand side should be larger that its decay rate. Taking into account the definition of the parameters, Eqs. (\ref{eq:eta}), (\ref{eq:r}) and (\ref{eq:A}), one comes to the generation threshold condition
\begin{equation}\label{eq:thres}
    4\pi\alpha^2\frac{d^2\sigma_m}{e^2c}f\mid\mu\mid y_{\pm}(qd) >1.
\end{equation}
where
\begin{equation}
    y_{\pm}(x)=\frac{\sin x}{x}\left(\frac{\sin x}{x}\pm 1\right)
\end{equation}
and $\alpha=e^2/(\hbar c)$ is the fine-structure constant. The functions $y_{\pm}(x)$ have positive maxima at $x\approx 1.5\pi,2.5\pi,3.5\pi,4.5\pi,...$, besides $x=0$.\\
Let us assume now that the driving field $E_0(t)$ has a broad frequency spectrum. Then the generation criterion can be in principle fulfilled for a range of frequencies. In such a case the situation resembles lasing in a multi-mode laser where only the mode
which has the largest excitation rate grows, while the other modes are discriminated.\cite{Haken} Therefore in what follows we shall only consider the mode which corresponds to the maximum self-excitation rate in Eq. (\ref{eq:thres}) and hence the minimum generation threshold. The smallest size of an oscillator is realized when $qd\approx 1.5\pi$ or, equivalently, $d/\lambda\approx 0.75/\epsilon^{1/2}$ with $\lambda$ the wavelength of the generated field. For this mode the criterion (\ref{eq:thres}) is reduced to
\begin{equation}\label{eq:thres1}
    \frac{3.2\alpha^2}{\epsilon^{1/2}}\frac{d^2\sigma_m}{e^2c}f\mid\mu\mid >1
\end{equation}
In particular, for the frequency of generation $\nu=\omega/(2\pi)=1$ THz and $\epsilon^{1/2}=1.5$ one obtains $d=150$ $\mu$m. Assuming for an estimate that the nanocomposite is loaded with silver nanorods  for which $\sigma_m=1.22\times 10^7$ S/m,\cite{Bordo23} one comes to the criterion of generation $f\mid\mu\mid>1.5\times 10^{-6}$ eV. It depends on both the volume fraction of nanorods and the chemical potential of the graphene sheets. For example, taking here $f=5\times 10^{-6}$ one gets $\mid\mu\mid>0.3$ eV that does not violate the adopted above condition $\mid\mu\mid\gg k_BT$ at room temperature.\\
In the course of the self-excitation process the electromagnetic field in the structure rapidly grows and so does the electric current induced in the metal nanorods that is accompanied by their Joule heating. As a result, the nanorod conductivity $\sigma_m$ decreases and when the inequality (\ref{eq:thres1}) is no longer satisfied, the process comes to the steady-state regime.\\
The self-excitation criterion, Eq. (\ref{eq:thres1}), can be rewritten in the form
\begin{equation}
    \frac{\sigma_m}{\sigma_{ss}}>1,
\end{equation}
where the quantity
\begin{equation}
    \sigma_{ss}=\frac{\epsilon^{1/2}}{3.2f\alpha^2}\frac{e^2c}{d^2\mid\mu\mid}
\end{equation}
can be called the steady-state conductivity. It indirectly determines the temperature rise, $\Delta T$, in the course of self-excitation and the attainable electric current in the nanorods, $I$. \\
The experimentally determined temperature dependence of the silver nanorod resistivity is given by\cite{Cheng15}
\begin{eqnarray}
    \rho(T)=\sigma_m^{-1}(T)\nonumber\\
    =3.25\times 10^{-8} \Omega\cdot\text{m}+1.68\times 10^{-10} \Omega\cdot \text{m}/\text{K}\times T.
\end{eqnarray}
Let us estimate the maximum attainable electric current in the nanorods, $I_{max}$. The increase in the resistivity is limited by the 
temperature $t_d\approx 400^{\circ}$ C above which the polymer matrix undergoes thermal degradation.\cite{Straus59} The maximum possible temperature rise above room temperature is therefore $(\Delta T)_{max}=380$ K. On the other hand, the temperature increase is related with the current through the relation\cite{Cheng15}
\begin{equation}
    \Delta T=\frac{I^2RL}{12kS},
\end{equation}
where $R=L/(\sigma_mS)$, $L$ and $S=\pi D^2/4$ are the nanorod resistance, length and cross section, respectively, with $D$ being the nanorod diameter, and $k=191.5$ W/(K$\cdot$m) is its thermal conductivity. Taking for an estimate $D=227$ nm and $L=27$ $\mu$m\cite{Cheng15} one obtains $I_{max}=7.7$ mA. \\
The value of $(\Delta T)_{max}$ dictates the lower limit for $\sigma_{ss}$, $\sigma_{ss}^{min}=6.87\times 10^6$ S/m that corresponds to $f\mid\mu\mid=2.7\times 10^{-6}$ eV. Accordingly, the maximum attainable electric field amplitude is given by $E_{max}=I_{max}/(\sigma_{ss}^{min}S)=280$ V/cm. This field should be compared with the saturation field in graphene, $E_s$, above which its optical response becomes nonlinear. For intraband transitions, $E_s=\omega E_F/(e\upsilon_F)$ with $E_F$ and $\upsilon_F$ the Fermi energy and the Fermi velocity in graphene, respectively.\cite{Marini17} Taking $E_F\sim 0.2$ eV and $\upsilon_F\sim 10^6$ m/s one obtains $E_s\sim 1.3\times 10^4$ V/cm, i.e. much larger than the field attainable in the oscillator. The field $E_{max}$ is also far below the breakdown strength typical for nanocomposite capacitors.\cite{Bordo22}\\
The power radiated by a single short antenna ($L\ll\lambda$) supporting an electric current $I$ is found as\cite{Antennas}
\begin{equation}
    \mathcal{P}_1=\frac{\pi L^2I^2}{3\epsilon_0c\lambda^2},
\end{equation}
where $\epsilon_0$ is the permittivity of free space. Multiplying it by the number of nanorods, 
\begin{equation}
 \mathcal{N}=\frac{f\mathcal{V}}{\mathcal{V}_1}, 
\end{equation}
where $\mathcal{V}=\mathcal{S}d$ is the oscillator volume with $\mathcal{S}$ its surface area and $\mathcal{V}_1$ is the nanorod volume,
one finds the total radiated power. Taking for an estimate $f=5\times 10^{-6}$, $\mathcal{S}=1\text{mm}\times 1\text{mm}$ and the other parameters as before one obtains $\mathcal{P}_{max}\approx 15$ mW that is quite sufficient for applications.\cite{Lu06} Actually, the total power can be even larger because the nanorods confined within a subwavelength nanocomposite film will radiate coherently in the direction perpendicular to the film.\\
The oscillator proposed in this paper can be fully controlled electrically. As a source of the driving (seed) terahertz field one can use just a wire incorporated in the structure and heated by an electric current. The generation threshold above which the self-excitation process develops is determined by the chemical potential of the graphene sheets which is dictated by their electrostatic gating. The generation can be triggered by applying the gating voltage which is above the threshold. On the other hand, this voltage can be switched off when necessary to interrupt the oscillator radiation, thus forming a terahertz pulse of desired duration. The wavelength of the generated radiation can be designed by a proper choice of the thickness of the cavity between the graphene mirrors. Besides that, such an oscillator would emit coherent radiation perpendicularly to the nanocomposite film, while the alignment of nanorods would ensure its polarization. In these respects, the proposed oscillator resembles a laser and it could find numerous applications in terahertz electronics.
\section*{Author Declarations}
\subsection*{Conflict of interest}
The author has no conflicts to disclose.
\section*{Data Availability}
The data that support the findings of this study are available from the corresponding author upon reasonable request.
\bibliography{tera}

\end{document}